\documentclass[electronic]{vgtc} 

\usepackage{tabu}      
\usepackage{booktabs}
\graphicspath{{./figs/}} 

\usepackage{times} 
\usepackage{amsfonts}
\usepackage{amsmath}
\usepackage{multirow}
\usepackage{enumitem}

\newcommand{\para}[1]{\vspace{0mm}\noindent{\textbf{#1}}}

\newcommand{\mm}[1] {\ifmmode{#1}\else{\mbox{\(#1\)}}\fi}

\newcommand{\Mspace}{\mathbb{M}}
\newcommand{\Rspace}{\mathbb{R}}

\newcommand{\Nyx}   {\mm{\mathsf{Nyx}}}
\newcommand{\microns} {\mm{\mathsf{MICrONS}}}
\newcommand{\etal}{et~al.}

\onlineid{1018}
\vgtccategory{Research}
\vgtcinsertpkg

\title{Extremely Scalable Distributed Computation of Contour Trees via Pre-Simplification}

\author{Mingzhe Li\thanks{e-mail: mingzhe.li@utah.edu}\\ %
        \scriptsize University of Utah %
\and Hamish Carr\thanks{e-mail: h.carr@leeds.ac.uk}\\ %
     \scriptsize University of Leeds %
\and Oliver R{\"u}bel\thanks{e-mail: oruebel@lbl.gov}\\ %
     \scriptsize Lawrence Berkeley National Laboratory %
\and Bei Wang\thanks{e-mail: beiwang@sci.utah.edu}\\
     \scriptsize University of Utah%
\and Gunther H. Weber\thanks{e-mail: ghweber@lbl.gov}\\
     \scriptsize Lawrence Berkeley National Laboratory
}

\abstract{
    Contour trees offer an abstract representation of the level set topology in scalar fields and are widely used in topological data analysis and visualization. However, applying contour trees to large-scale scientific datasets remains challenging due to scalability limitations. Recent developments in distributed hierarchical contour trees have addressed these challenges by enabling scalable computation across distributed systems. Building on these structures, advanced analytical tasks—such as volumetric branch decomposition and contour extraction—have been introduced to facilitate large-scale scientific analysis. Despite these advancements, such analytical tasks substantially increase memory usage, which hampers scalability. In this paper, we propose a pre-simplification strategy to significantly reduce the memory overhead associated with analytical tasks on distributed hierarchical contour trees. We demonstrate enhanced scalability through strong scaling experiments, constructing the largest known contour tree—comprising over half a trillion nodes with complex topology—in under 15 minutes on a dataset containing 550 billion elements.

} 

\keywords{Contour Tree, Computational Topology, Distributed Algorithm, Branch Decomposition, Topological Data Analysis}

\begin{document}
\maketitle

\section{Introduction}
\label{sec:introduction}

Scientific and engineering simulations have long been cornerstones of supercomputing. 
However, as the scale and complexity of data have continued to expand, the human capacity to assimilate and comprehend such data has remained relatively static. 
As the bandwidth of the human visual system remains inherently limited, the need for advanced data analysis and visualization tools has become increasingly pressing. 
One of the most successful tools in this domain to date has been computational topology, which leverages rigorous mathematical frameworks to enhance the visualization and interpretation of complex data. As topology-based analysis and visualization techniques have matured, attention has increasingly turned toward their parallelization and distribution—key steps required to fully realize their potential at the \emph{exascale} where they are most critically needed. 

One important topological descriptor is the contour tree, which analyzes contours (level sets of scalar fields) to reveal relationships between features defined by critical points, enables hierarchical simplification based on geometric properties, and serves as a foundational structure for isosurface-based visualization techniques.

Recent work under the ECP ALPINE project~\cite{AAA25} introduced efficient parallel algorithms within the PRAM model (Parallel Random Access Machine) for contour tree construction~\cite{CarrWeberSewell2016,CarrWeberSewell2021}, and methods for augmenting the contour tree and accelerating access to its structure~\cite{CarrRubelWeber2022a}. Advances have also been made in computing geometric properties, performing simplification, and supporting visualization~\cite{HristovWeberCarr2020}. 
Building upon efficient on-node parallelism, subsequent work has achieved additional performance gains by leveraging distributed parallelism—both for contour tree computation~\cite{CarrRubelWeber2022b} and for tasks such as augmentation, geometric property computation, and visualization~\cite{LCR24}.

Despite its strengths, contour tree computation has constraints that can limit the effectiveness of even the most advanced algorithms. 
First, the memory footprint can be $100\sim200$ bytes per cell for regular meshes, whose intrinsic format stores neighborhood information implicitly.  
Second, because topological analysis derives inherently global properties, substantial data exchange is needed between machines in a cluster, leading to bottlenecks in both inter-node communication and per-node storage.  
As a result, although the most recent work~\cite{LCR24} achieved up to a $100\times$ speedup in distributed settings—on top of prior single-node gains of up to $200\times$—the full analytic pipeline became impractical for data sizes beyond $1024^3$. 
This was the case even though the core contour tree computation~\cite{CarrRubelWeber2022b} successfully handled volumes as large as $2048 \times 2048 \times 4096$  on a previous-generation system, and a related but simpler computation of merge trees scaled to $8192^3$ data~\cite{NigmetovMorozov2019}.

\para{Contribution.}~In this paper, we apply a novel pre-simplification strategy that substantially reduces the memory overhead associated with analytical tasks on distributed hierarchical contour trees, thereby enabling highly scalable computation.
For simulation data, the final stages of analysis and visualization typically rely on contour tree simplification to separate features from noise. 
We enhance the existing implementation by performing simplification both \emph{before} and \emph{after} the construction of the distributed contour tree, thereby reducing per-node memory footprint as well as inter-node communication cost.

Our framework enables contour tree analysis of the largest dataset reported to date—$8192^3$ grid points, or approximately 550 billion elements, corresponding to 4 TiB (tebibytes) of data—in less than 15 minutes using approximately 120 TiB of working memory distributed across 512 nodes. 

\para{Overview.}~We review contour tree computation in \cref{sec:background} and introduce pre-simplification for improved scalability in \cref{sec:method}. Implementation details are provided in \cref{sec:implementation}, where our code is integrated into VTK-m~\cite{MorelandSewellUsher2016} (recently renamed to Viskores). Results are presented in \cref{sec:results}, and conclusions are given in \cref{sec:conclusion}.

\begin{figure*}[!ht]
\centering
\includegraphics[width=0.95\linewidth]{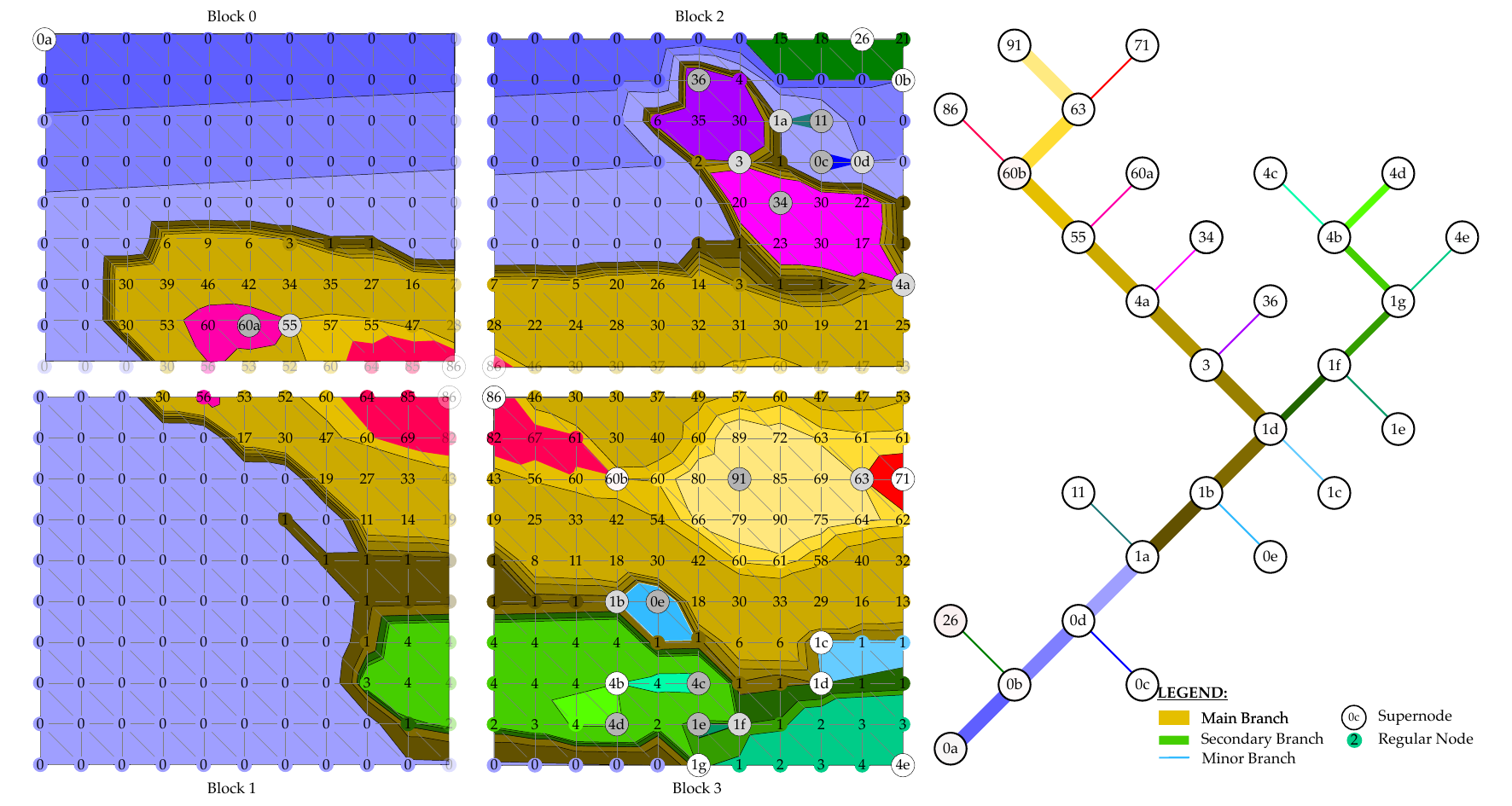}
\vspace{-4mm}
\caption{Contour tree of a small dataset for the elevation of Vancouver, color-coded by topological zone/feature, with branch decomposition by area measure approximated with vertex count. Lettered values indicate the sorting order under simulation of simplicity~\cite{EdelsbrunnerMucke1990}. 
Figure adapted from Li et al.~\cite[Fig.~2]{LCR24} \copyright~IEEE, with permission.}
\vspace{-4mm}
\label{fig:vancouver}
\end{figure*}

\section{Background}
\label{sec:background}

Large-scale simulations investigate physical phenomena by numerically modeling physical properties, typically as continuous functions defined over a spatial domain. This domain is usually discretized into a mesh composed of individual cells, most commonly tetrahedral or cubic in shape. 
In this paper, we assume the data of interest is given as a function $f: \Mspace \to \Rspace$ (where the manifold $\Mspace$ is a subset of $\Rspace^2$ or $\Rspace^3$) with a suitable discretization. 

We start by introducing the contour tree and its utilization in visualization (\cref{sec:ct}). 
We then introduce the computation of contour tree via serial (\cref{sec:serial-ct}), locally parallel  (\cref{sec:parallel-ct}), and distributed algorithms (\cref{sec:distributed-ct}).  

\subsection{Contour Trees}
\label{sec:ct}

Given a function $f: \Mspace \rightarrow \Rspace$, the \emph{level set} $f^{-1}(a)$ defines a subset of points in $\Mspace$ at a given \emph{isovalue} $a$; these are called \emph{isolines} in 2D and \emph{isosurfaces} in 3D. Each non-empty level set contains one or more connected components called \emph{contours}, and we can study the function $f$ by analyzing how these contours change as the isovalue $a$ varies. 
Two points $x, y \in \Mspace$ are considered \emph{equivalent}, denoted $x \sim y$, if they have the same isovalue and belong to the same contour.  
Based on such an equivalent relation, we contract each contour to a single point, giving rise to a quotient space $\Mspace/\sim$ first described by Reeb~\cite{Reeb1946}. 
This quotient space collapses the data $(\Mspace, f)$ to a skeletal structure---called the \emph{Reeb graph}---which connects local minima, local maxima, and saddles of $f$.

If a Reeb graph contains no cycles (i.e.,~when the domain $\Mspace$ is simply connected), it is a \emph{contour tree}. The contour tree was originally defined independently as a data structure for efficient access to polygonal contours stored in memory~\cite{BoyellRuston1963}. 
We show a small example of a 2D contour tree in \cref{fig:vancouver}.

In a contour tree, \emph{supernodes} occur at critical points of $f$ (points with zero gradient), which are connected by \emph{superarcs} representing equivalence classes of contours that map to \emph{topological zones} in the input domain. 
Critical points where the number of contours does not change are not supernodes, for example, when an isosurface changes its genus. 
Most algorithms in computing contour trees begin by operating on simplicial meshes, where critical points are guaranteed to lie at the mesh vertices~\cite{Ban67}. 

A contour tree can be \emph{augmented} with \emph{regular nodes}—most commonly the mesh vertices that are not critical points—as these nodes are essential for computing the geometric properties of the associated topological zones. 
If needed, these regular nodes are connected by \emph{regular arcs} that break up the parent superarcs.  
Analysis of contour tree algorithms therefore depends primarily on the data size $n$ (normally the same as the number of mesh vertices), sometimes the number of edges in the mesh $m$ (which is $O(n)$ for regular meshes but may be $O(n^2)$ for irregular meshes), and the number of supernodes $t$ in the contour tree.

The contour tree is used for terrain~\cite{BoyellRuston1963,FM67,GC86}, to accelerate isosurface extraction~\cite{IK94,KreveldOostrumBajaj1997,CS03}, and for feature extraction~\cite{CarrSnoeyinkPanne2010}, with variations adapted for volume rendering~\cite{WeberDillardCarr2007}, protein molecule comparison~\cite{ZhangBajajBaker2004}, and energy landscape analysis~\cite{HW10}.

\subsection{Serial Contour Tree Computation}
\label{sec:serial-ct}

van Kreveld~\etal~\cite{KreveldOostrumBajaj1997} reported a serial algorithm for computing a contour tree. 
The algorithm performs a sweep of a contour through the simplicial mesh from high to low isovalues, explicitly tracking the cells intersected by the contour and updating this set as the sweep progresses through each mesh vertex. The algorithm has a time complexity of $O(m \log m)$ in 2D and $O(m^2)$ in higher dimensions. 
Tarasov and Vyalyi~\cite{TV98} extended this framework to $O(m \log m)$ in 3D with a complex three-sweep algorithm that requires special processing at the boundaries as well as repeated subdivision of the mesh to avoid topological problems.

Carr~\etal~\cite{CarrSnoeyinkAxen2003} gave a serial algorithm that sweeps downwards through the data to identify relationships between local maxima and saddles, then upwards for local minima and saddles, producing two \emph{merge trees}. Leaf edges are then transferred serially from the merge trees to construct the contour tree recursively.  With an initial sorting and a modified union-find algorithm, the overall cost is $O(n \log n + \alpha(m,n))$, where $\alpha$ is the inverse Ackermann function. Later work~\cite{PascucciColeMcLaughlin2003} relaxed the initial assumption of a simplicial mesh, allowing trilinear interpolation on cubical cells. 
A more general approach~\cite{CarrSnoeyink2009} identified that the algorithm could use a \emph{topology graph} constructed for arbitrary meshes and interpolants, such as the approximate trilinear interpolation from Marching Cube isosurfaces.

\begin{figure*}[!ht]
\centering
\includegraphics[width=1.0\linewidth]{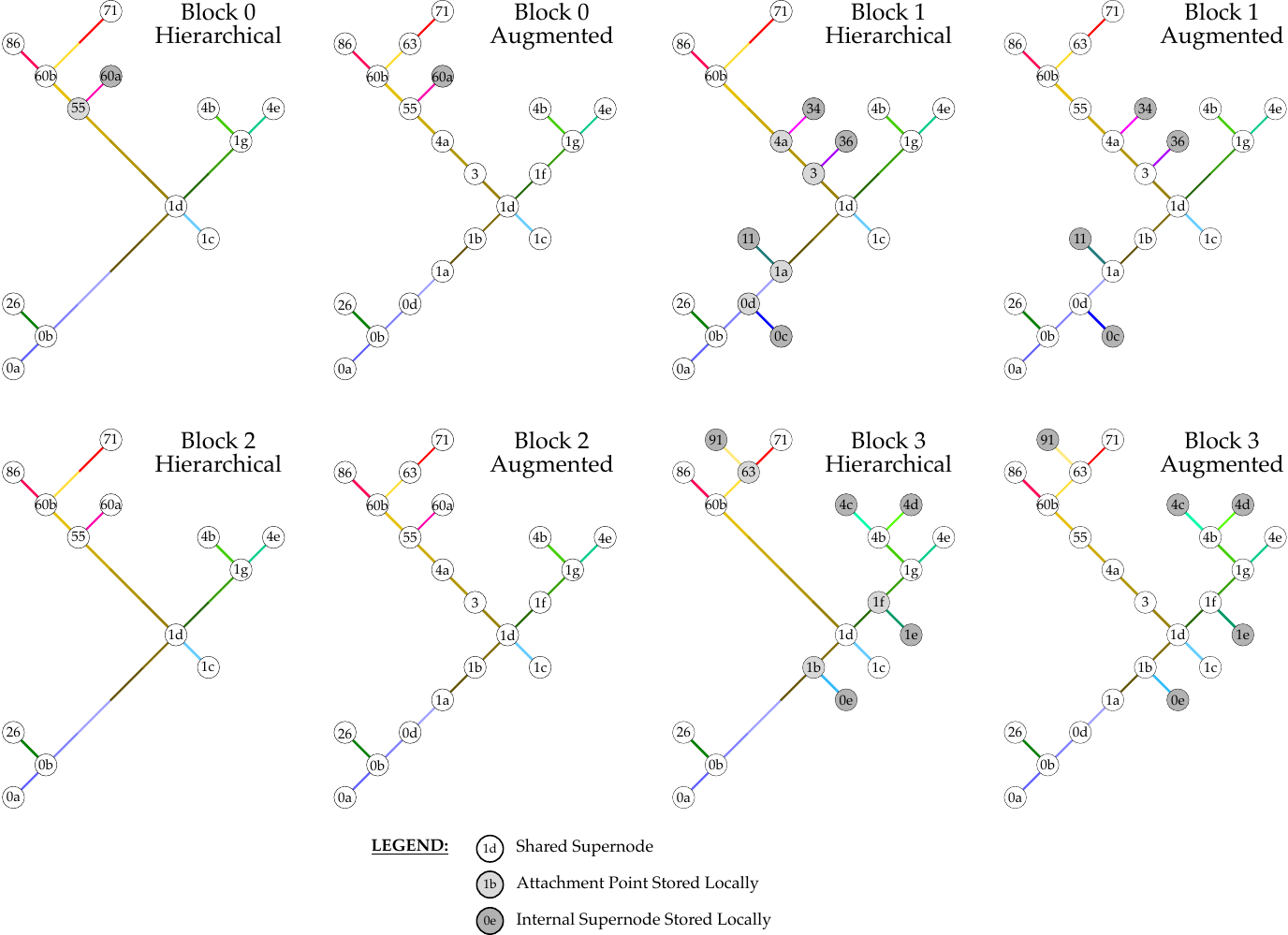}
\vspace{-6mm}
\caption{Distributed hierarchical contour tree for \cref{fig:vancouver}, augmented for volume computation and branch decomposition. Augmentation increases the number of shared vertices, which becomes a bottleneck for distributed storage and communication. 
Figure adapted from Li et al.~\cite[Fig.~3]{LCR24} \copyright~IEEE, with permission.}
\vspace{-6mm}
\label{fig:dhct}
\end{figure*}

Computing the contour tree alone is insufficient for many analytical tasks such as branch decomposition, contour tree simplification, and contour extraction.  
For instance, contour tree simplification recursively discards leaves representing insignificant features to produce a \emph{branch decomposition}~\cite{PascucciColeMcLaughlin2003}. 
This process is different from the cancellation of critical point pairs~\cite{ELZ00}, when complex structures (so-called \emph{W-structures}~\cite{HristovCarr2021}) are present in the tree. 

Geometric properties—such as contour surface area and enclosed volume—are also useful for identifying significant isosurfaces~\cite{BajajPascucciSchikore1997}. 
Subsequent work computed these properties for individual contours by sweeping inward through the contour tree~\cite{CarrSnoeyinkPanne2010}, enabling their use as measures of importance for simplification. \cref{fig:vancouver} shows  the decomposition of the contour tree into branches based on surface area, with more important branches shown as thicker lines.

While branches are often thought of as linear paths within the contour tree, it is better to think of them as subtrees anchored to a single parent branch or trunk. This is because a contour taken just above the attachment point—where the branch connects to its parent—encloses all regions within the corresponding subtree. For instance, in \cref{fig:vancouver}, the branch \(4d\text{--}1d\) connects to its parent at \(1d\), and the contour at \(1d + \epsilon\) (for an arbitrarily small $\epsilon$) encloses not only the sequential regions \(1d\text{--}1f\), \(1f\text{--}1g\), \(1g\text{--}4b\), and \(4b\text{--}4d\) along the branch itself, but also the child branches \(1g\text{--}4e\) and \(4b\text{--}4c\).

\subsection{Parallel Contour Tree Computation}
\label{sec:parallel-ct}

Although a distributed parallel algorithm was introduced relatively early~\cite{PascucciColeMcLaughlin2003}, the development of shared-memory parallel algorithms for contour trees emerged only after a significant delay. 
Acharya and Natarajan~\cite{AcharyaNatarajan2015} used GPU parallelism to build monotone paths between critical points, then used them as a topology graph for the serial algorithm on CPU.
Carr~\etal~\cite{CSL15} gave a distributed algorithm for computing contour trees on GPU based on quantizing the mesh, then contracting all contours in parallel with union-find. 
However, this approach was difficult to validate and incurred a substantial memory footprint, significantly limiting its practical utility.

Gueunet~\etal~\cite{GueunetFortinJomier2016,GueunetFortinJomier2017} segmented the mesh by isovalues, then computed separate contour trees for each segment on individual threads and merged them with a task-based approach, achieving peak speedups of approximately $10\times$. Smirnov and Morozov~\cite{SmirnovMorozov2020} used a task-based algorithm to add mesh edges incrementally, collapsing redundant edges to build merge trees.
Carr~\etal~\cite{CarrWeberSewell2016} presented a PRAM algorithm that employs pointer-jumping to construct monotone paths, which are then used to compute a topology graph. 
The algorithm then identifies in parallel one superarc from each remaining extremum to its governing saddle—i.e. the last saddle at which a monotone path can reach another extremum instead.  Removing each extremum from the topology graph and redirecting its paths to its governing saddle then constructs a smaller topology graph in which some or
all of the saddles become extrema.  After a logarithmic number of passes, all superarcs in the merge tree have been identified.

A second phase replaced recursive leaf transfer from the merge trees with batched alternating transfers of all upper or lower leaves, using pointer-doubling to collapse vertical chains of vertices left by removing leaves in each pass. This algorithm achieves a polylogarithmic computational cost and demonstrates practical speedups of nearly $50\times$ over the standard serial implementation.

Later work~\cite{CarrRubelWeber2022a} added an accelerating \emph{hyperstructure} related to rake-and-contract~\cite{MR89} to provide logarithmic access into the tree, allowing efficient augmentation with regular nodes. Each \emph{hyperarc} in this hyperstructure captures a vertical chain collapsed in each pass of the batched transfer, and stores the superarcs (and supernodes) on each hyperarc in sorted order, permitting binary search by data value for rapid access into the tree.

Hristov~\etal~\cite{HristovWeberCarr2020} employed segmented prefix-scan operations to sum values along the hyperarcs, replacing serial sweeps through the tree with parallelized \emph{hypersweeps} over the hyperarcs to compute geometric properties. They then extracted a branch decomposition by using approximated volume as the measure, selecting the “most important” up and down arcs at each vertex in parallel, and applying pointer-jumping to collapse the branches.

Analytically, these algorithms run polylogarithmically in time, except where W-structures are present in the contour tree~\cite{HristovCarr2021}; however, the impact of these structures is minimal in practice. 

\subsection{Distributed Contour Trees}
\label{sec:distributed-ct}

Pascucci and Cole-McLaughlin~\cite{PascucciColeMcLaughlin2003} presented a distributed algorithm that computes individual contour trees for blocks of a dataset and then unites them into a topology graph for the combined contour tree. While effective in principle, the approach computes and stores the final contour tree on a single node. Given the substantial memory footprint discussed earlier, this limitation renders the algorithm ineffective for distributed computation until a fully distributed data structure is developed.

Nigmetov and Morozov~\cite{NigmetovMorozov2019} modified a previous local task-based parallel approach~\cite{MorozovWeber2013} with a triplet merge tree representation~\cite{SmirnovMorozov2020} to compute a distributed representation of the merge tree. 
Landge~\etal~\cite{LPG14} gave an algorithm that discards local features of merge tree, combining the remainder incrementatlly using a fan-in process. 
However, these approaches did not exploit local parallelism, and only computed the merge trees, without any of the secondary geometric or topological properties. 

Carr~\etal~\cite{CarrRubelWeber2022b} adapted the previous distributed approach~\cite{PascucciColeMcLaughlin2003} by removing \emph{interior forests} of superarcs that only exist on a given block from the contour tree at each stage of a fan-in process.  The interior forests are then reinserted during a fan-out phase to build a \emph{distributed hierarchical contour tree} (DHCT), which distributes shared superarcs across multiple machines, in a layered version of the hyperstructure.  
Peak speedups were reported of $70\times$ compared to the local PRAM algorithm, and maximum data size on NERSC's Cori supercomputer was $2048 \times 2048 \times 4096$.

Recently, Li~\etal~\cite{LCR24} observed that certain geometric computations depend on the precise ordering of superarcs along a hyperarc. Because the distribution strategy for the contour tree treated critical points with only local significance as regular points during construction, they introduced an additional stage to augment the tree with these points before implementing a distributed version of the hypersweep and simplification; see \cref{fig:dhct} for augmented contour trees corresponding to the example in \cref{fig:vancouver}. While this approach improved performance, the extra memory and communication required for all machines to process these supplementary \emph{attachment points} limited the largest dataset computed and analyzed on NERSC’s Perlmutter supercomputer to $1024^3$.

\section{Method}
\label{sec:method}

In the work of Li et al.~\cite{LCR24}, the contour tree computation itself scaled well, but secondary computations—such as geometric property evaluation and simplification—did not, due to the memory overhead of augmenting the local data structure with attachment points for interior forests on other machines. In this paper, to restore scalability following~\cite{LCR24}, we focus on analytic tasks with distributed contour trees, namely simplification and extraction.

The primary goal of the analysis is to extract a manageable set of significant contours that capture the key features of the data. This is commonly accomplished by simplifying the contour tree to a fixed number $b$ of branches or by applying a fixed threshold $\Lambda$ on an importance measure, which may—but need not—coincide with the measure originally used for branch definition. For clarity, we assume volume is used as the criterion for both branch decomposition and simplification, with a predefined volume threshold delineating the features of interest. 

Since only a small number of important contours are selected, and most augmenting vertices serve as attachment points for small branches or subtrees, we simplify as many of these subtrees as possible \emph{prior} to tree construction. This reduces the number of augmenting vertices and, in turn, lowers both memory footprint and communication cost. We introduce an additional threshold $\lambda$ to designate subtrees as sufficiently unimportant to exclude from communication with neighboring nodes. We refer to this step as \emph{pre-simplification}.

If $\lambda \leq \Lambda$, pre-simplification does not affect the final selection of contours for measures such as volume, which increase monotonically as one sweeps through the tree. Pre-simplifying a given subtree is equivalent~\cite{CarrSnoeyinkPanne2010} to replacing the function value throughout the subtree’s topological zone with the value of the saddle point—or, in the case of a simplicial mesh, setting the values of all regular nodes in that zone to the saddle value. Consequently, we can treat all such nodes as belonging to the same superarc as the attachment point, ensuring that their contribution to the parent subtree’s measure remains unaffected by pre-simplification. As a result, any subtree represented in the pre-simplified contour tree will retain the correct volume, and the desired outcome follows.

\begin{figure}[!ht]
\centering
\includegraphics[width=1.0\columnwidth]{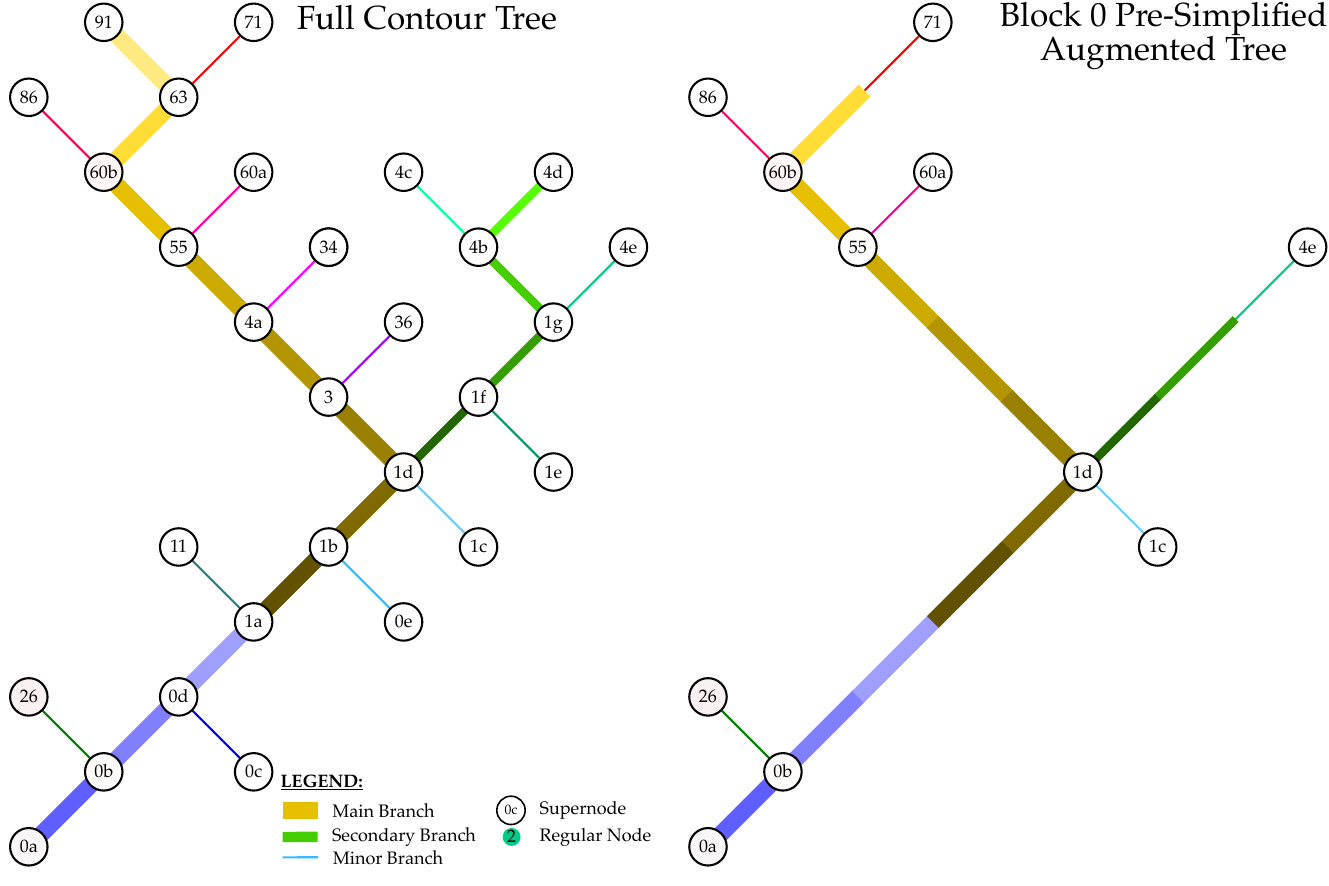}
\vspace{-6mm}
\caption{A branch with the wrong leaf. Vertex $4d$ lies within the interior of a block and can thus be pre-simplified, whereas vertex $4e$, located on the block's boundary, cannot. After pre-simplification and augmentation, a branch will be selected from $1d-4e$ rather than $1d-4d$. Nonetheless, the subtree rooted at $1d$ is accurately represented.}
\label{fig:wrong-leaf}
\end{figure}

One side effect is that, although remaining subtrees maintain the correct volume, the branch representing the subtree may have a different outer leaf in the pre-simplified contour tree. To understand why this occurs, consider \cref{fig:wrong-leaf}. In this case, vertex \(4d\) would typically be selected as the outer leaf of a branch rooted at \(1d\), presenting the subtree containing \(1f\), \(1e\), \(1g\), \(4e\), \(4b\), \(4c\), and \(4d\). However, since \(4d\) lies within the interior of a block and \(4e\) cannot serve as an attachment point due to its position on the boundary, the algorithm retains \(4e\) and reports the branch \(1d-4e\) instead.

We outline the high-level steps for pre-simplified distributed hierarchical contour tree computation and branch decomposition, adapted from the pipeline in~\cite{LCR24}:
\begin{enumerate}[noitemsep,leftmargin=*]
\item Compute a distributed hierarchical contour tree;
\item Compute the correct volume for all superarcs in the tree;
\item For each block, list attachment points with measure $> \lambda$;
\item Augment the distributed tree with these attachment points;
\item Recompute the volumes for all superarcs in the tree;
\item Compute the distributed branch decomposition;
\item Simplify the contour tree to threshold $\Lambda$ or to $b$ branches;
\item Extract and render the contours.
\end{enumerate}
Steps 1 and 2 remain unchanged, while Steps 7 and 8 are only minimally affected. The remaining steps require further explanation.

Step 3 (listing attachment points) identifies the subtrees of the interior forest with measure $> \lambda$ in a single logarithmic-cost PRAM pass. At this stage, the internal logic must be substantially revised to support augmentation~\cite{LCR24}. The original goal in \cite{LCR24} was to avoid explicitly representing the supernodes connecting interior trees, thereby reducing communication costs in both fan-in and fan-out operations. However, situations arise—such as at vertex $1d$ in \cref{fig:vancouver}—where insertions occur at multiple levels of the hierarchy.

In the original DHCT, interior trees were stored by setting the superarc of the attachment point to null, as with the root in the non-distributed structure. A second array stored the \emph{superparent}—the superarc to which each regular node (including supernodes) belonged. For efficiency, each superarc was indexed by the ID of its outer end; thus, supernodes normally had their superparent set to their own ID.

For attachment points already in a higher layer, the superparent could be assumed correct. Otherwise, the superarc ID into which the attachment point was to be inserted was stored as the superparent, avoiding the need for an additional array.

In the augmented distributed tree~\cite{LCR24}, a new distributed structure was constructed rather than editing the existing one. All attachment points were inserted from the outset, so the distributed hypersweeps, branch decomposition, and simplification routines did not need to check both superarc and superparent. With the introduction of pre-simplification, however, these routines needed to handle the possibility of uninserted attachment points, complicating Steps 4 through 7.

Step 4 (augmentation) proceeded largely as before, but with only a subset of attachment points exchanged during fan-in. After augmentation, the hierarchical tree’s supernode IDs often differed, requiring recomputation of the volumes for each superarc and subtree (Step 5).

Steps 6, 7, and 8 were also modified to use new representative branch IDs. In the original approach~\cite{LCR24}, the correct extremum for each branch was known, and the branch ID was taken to be the outer end’s ID. The mislabeling introduced by pre-simplification rendered this strategy unreliable. Instead, the saddle at the root of the subtree was used as the representative, again necessitating substantial changes in the detailed processing.

\section{Implementation and Experiments}
\label{sec:implementation}

We begin with implementation details in~\cref{sec:implement-details}, followed by a description of the datasets in~\cref{sec:datasets}. We then present the  experimental setup in~\cref{sec:experiment-setting}, and conclude with a parameter sensitivity analysis of the pre-simplification threshold in~\cref{sec:parameter-sensitivity}.

\subsection{Implementation Details}
\label{sec:implement-details}
Our implementation is based on VTK-m~\cite{MorelandSewellUsher2016}, an open-source library for efficient scientific visualization algorithms enhanced with on-node SMP parallelism. 
For distributed computation, we use the DIY~\cite{morozov2016block} block-parallel library. 
We implement the pre-simplification strategy on top of the distributed hierarchical contour tree implementation in the VTK-m library.
Specifically, we modify the \texttt{ContourTreeUniformDistributed} filter to update the pipeline with the pre-simplification strategy in~\cref{sec:method}. 
We add a hypersweep computation to superarc volumes before the augmentation, referred to as the \emph{pre-augmentation hypersweep}.
We only augment the contour tree with attachment points whose corresponding interior forest volume is higher than $\lambda$.
In addition, we update the \texttt{SelectTopVolumeBranches} filter to limit branch selection in the simplified contour tree to branches with a volume above $\lambda$ (i.e., to ensure $\lambda \leq \Lambda$).
Our implementations are available at \href{https://github.com/Viskores/viskores}{https://github.com/Viskores/viskores}.

\subsection{Datasets}
\label{sec:datasets}

\begin{figure*}[!ht]
\centering
\includegraphics[width=2.0\columnwidth]{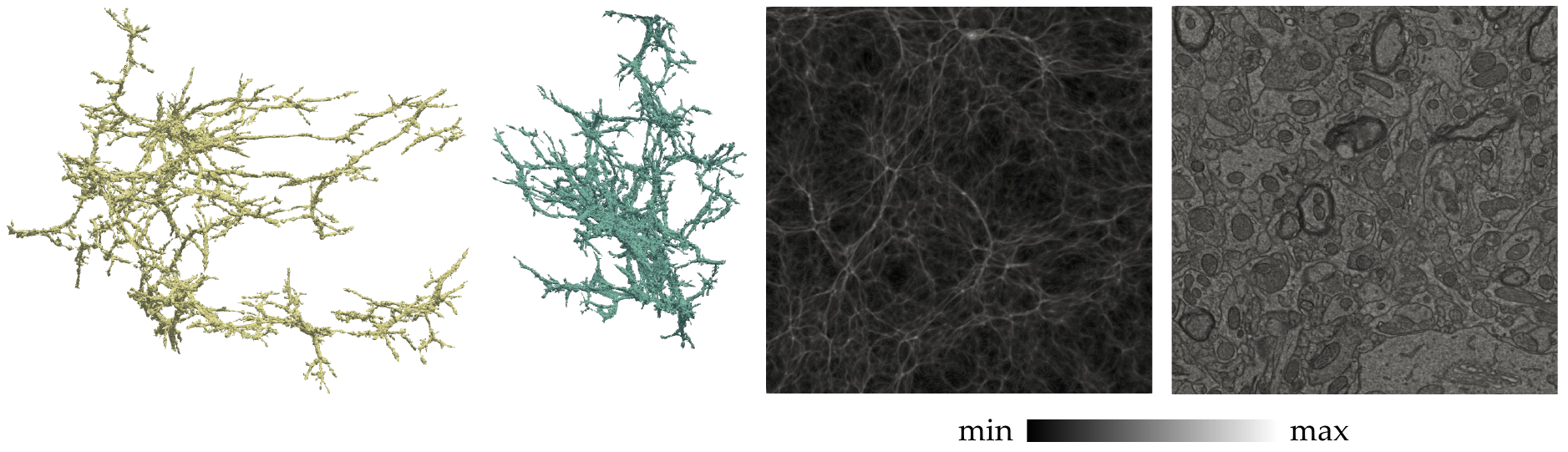}
\vspace{-4mm}
\caption{Visualizations of the {\Nyx} and {\microns} datasets using our framework. 
Left: two 3D contours of high-volume features extracted from a $1024^3$ subvolume of the {\Nyx} dataset. Right: 2D slices of the {\Nyx} dataset and the {\microns} dataset respectively.}
\label{fig:data-visualization}
\vspace{-3mm}
\end{figure*}

We experiment with two large datasets: {\Nyx} and {\microns}. 
{\Nyx} is a $4096^3$ dataset from cosmological simulations of particle mass density~\cite{AlmgrenBellLijewski2013}: we use matter density $\Omega_m$~\cite{lambdaNASA}) (the sum of baryon density and dark matter density) as the scalar field in a 3D volume.

{\microns}~\cite{ConsortiumBaeBaptiste2021} is a volume of Electron Microscopy (EM) image data of a P60 mouse cortex with a volume of $1.4mm \times 0.87mm \times 0.84mm$ from the BossDB~\cite{bossdb}. We work on a volume of $8192^3$ voxels cropped from the original data.

\cref{fig:data-visualization} showcases visualizations for both datasets. 
In the 3D visualization for the {\Nyx} dataset (1st and 2nd column), there are two 3D contours rendered in different colors, corresponding to the $68$th and $89$th highest volume branch of the contour tree.
These extracted contours highlight the filamentary structures of matter density in the universe.
The 3rd and 4th columns of~\cref{fig:data-visualization} show the visualizations for 2D slices of the {\Nyx} and the {\microns} dataset, respectively. 
In the 2D image of the {\Nyx} dataset, there are similar filamentary structures to the 3D contours. 
In contrast, the {\microns} image shows many cell-structure shapes in the mouse cortex.
The irregular and intricate shapes of features in both datasets contribute to a contour tree structure that is correspondingly large and complex.

\subsection{Experimental Settings}
\label{sec:experiment-setting}

\para{Hardware configurations.}
All experiments are conducted on the National Energy Research Scientific Computing Center (NERSC)'s Perlmutter supercomputer with $3,072$ CPU-only and $1,792$ GPU-accelerated nodes. Our experiments were conducted on CPU-only nodes, each with two 2.45 GHz (up to 3.5 GHz) AMD EPYC 7763 (Milan) CPUs with 64 cores per CPU and two hardware threads per physical core, and 512 GB of DDR4 memory per node.

\para{Computational parameter configurations.}
For all the experiments, we fix one data block per MPI rank and one MPI rank per CPU node. 
This is to reduce the number of volume subdivisions to minimize the size of boundary information that has to be shared across blocks, which leads to a scalability bottleneck in the distributed hierarchical contour tree framework~\cite{CarrRubelWeber2022b}. 
For each compute node, we use $128$ threads with OpenMP~\cite{dagum1998openmp} for thread parallelism. 

\para{Algorithmic configurations.}
We retain the $b$ branches with the largest volumes to simplify the contour tree based on the branch decomposition, fixing $b = 100$ for all experiments. Li \etal~\cite{LCR24} showed that the runtime of contour tree simplification is insensitive to $b$; we select a small value so that no branch retained in the simplified contour tree is subject to pre-simplification. Parameter sensitivity with respect to $\lambda$ is analyzed in~\cref{sec:parameter-sensitivity}. We omit contour extraction runtimes—reported in~\cite{LCR24}—as they depend primarily on contour size rather than contour tree structure, and thus their scalability lies outside the scope of this work.

\para{Runtime evaluation configuration.}
We report runtime of different pipeline phases: the first three phases, namely (1) computing the local contour tree, (2) fan-in, and (3) fan-out, contribute to the contour tree construction, which are the same as in the previous work~\cite{LCR24}.
The subsequent phases are the analytical tasks, including (4) the newly introduced pre-augmentation hypersweep, (5) augmentation, (6) post-augmentation hypersweep, (7) branch decomposition, and (8) extraction of the top-volume branches. Any remaining computational costs are grouped as ``Other.''

To collect runtime statistics, we place synchronization barriers after each phase. Due to the sequential nature of the pipeline, these barriers do not introduce significant overhead. For each phase, we report maximum runtime observed across all MPI ranks. Note that inter-rank communication occurs during the fan-in phase (Phase 2) and throughout the analytical computation stages (Phases 4–8).

\begin{figure}[!ht]
\centering
\includegraphics[width=1.0\columnwidth]{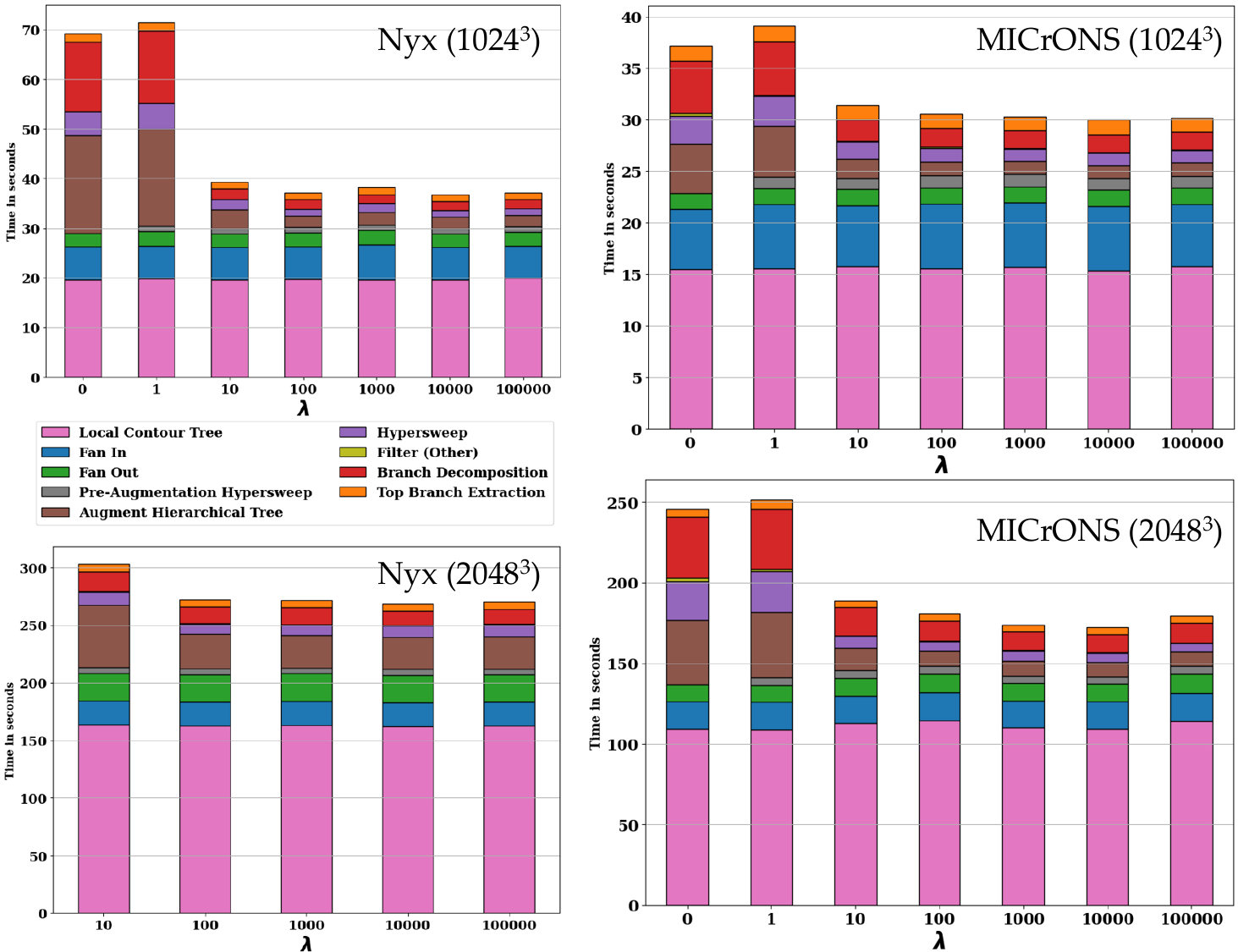}
\vspace{-3mm}
\caption{Overall runtime using 16 nodes for the $1024^3$ and $2048^3$ subvolumes of the {\Nyx} (left column) and the {\microns} (right column) datasets, respectively, with $\lambda$ ranging from $0$ to $10^5$.}
\label{fig:parameter-runtime}
% \vspace{-2mm}
\end{figure}

\begin{figure}[!ht]
\vspace{-2mm}
\centering
\includegraphics[width=1.0\columnwidth]{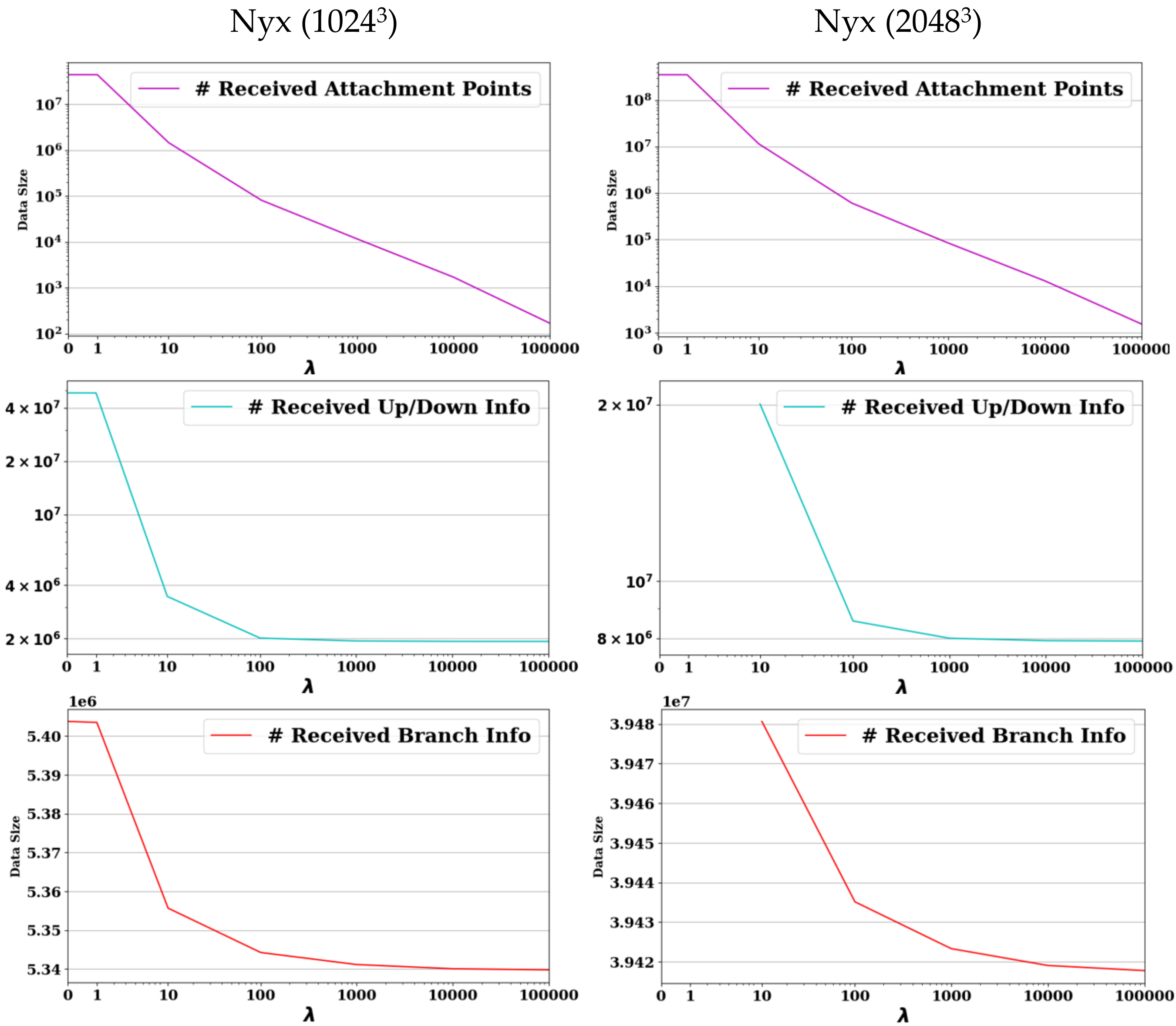}
\vspace{-3mm}
\caption{The communication workload of attachment points (1st row, log-log), best up/down volume information (2nd row, log-log), and the branch information (3rd row, log-linear) for the $1024^3$ and $2048^3$ subvolumes of the {\Nyx} dataset, respectively, with $\lambda$ between $0$ and $10^5$. 
Each statistic is collected on the rank with the highest workload.}
\label{fig:parameter-Nyx-communication}
% \vspace{-2mm}
\end{figure}

\begin{figure}[!ht]
% \vspace{-2mm}
\centering
\includegraphics[width=0.95\columnwidth]{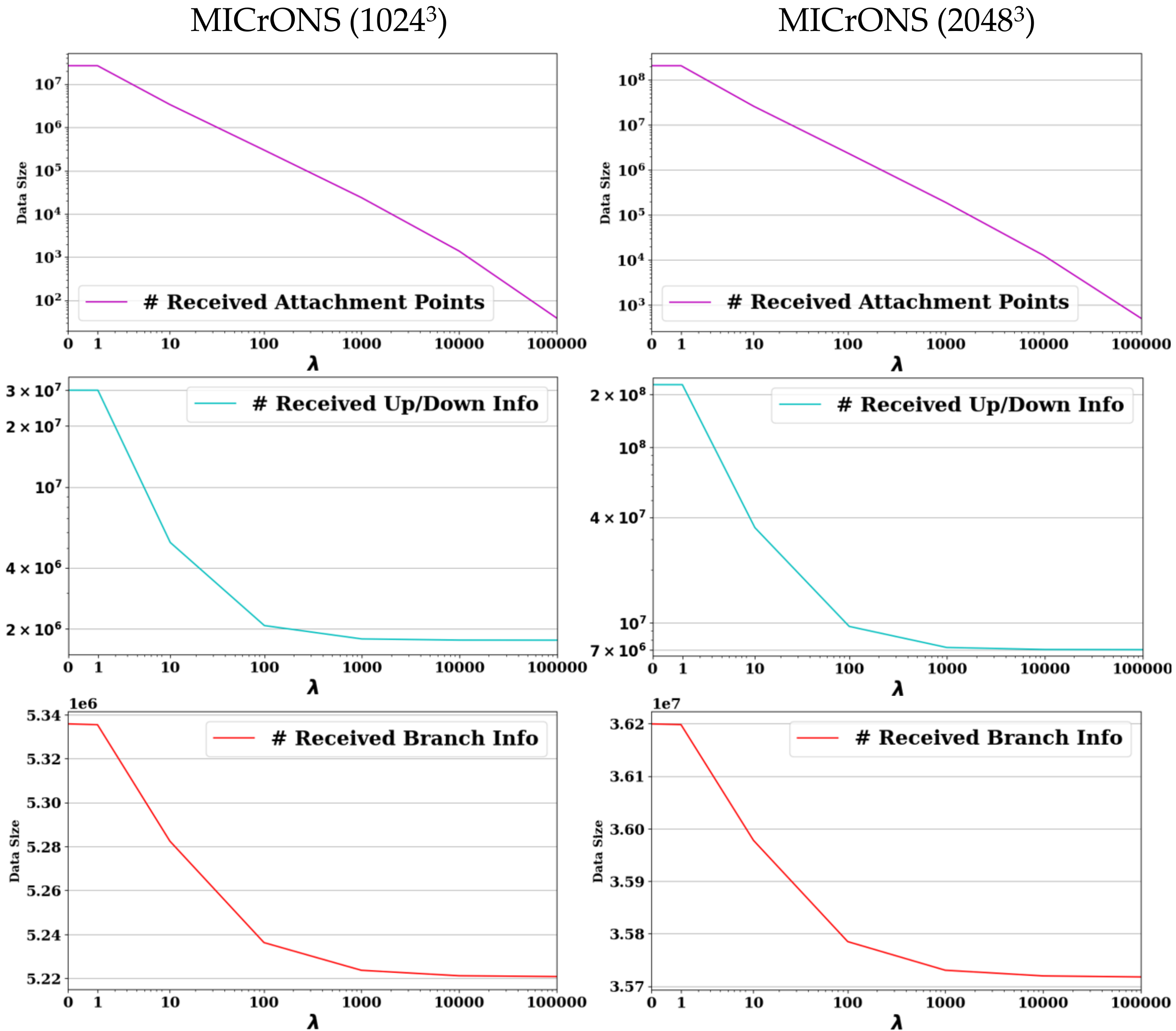}
\vspace{-2mm}
\caption{Communication workload of attachment points (1st row, log-log), best up/down volume information (2nd row, log-log), and branch information (3rd row, log-linear) for $1024^3$ and $2048^3$ subvolumes of {\microns}, with $\lambda$ ranging from $0$ to $10^5$. 
Each statistic is collected on the rank with the highest workload.}
\label{fig:parameter-microns-communication}
\vspace{-2mm}
\end{figure}

\subsection{Parameter Sensitivity Analysis}
\label{sec:parameter-sensitivity}

Recall that we only exchange attachment points whose subtree volumes are higher than $\lambda$ during augmentation, reducing the overall communication cost for augmentation and subsequent steps.
In theory, the correctness of the final selection and extraction of contours is guaranteed if $\lambda < \Lambda$, which is the volume for the branch with the $b$-th highest volume.
Here, we conduct a parameter sensitivity analysis for $\lambda$ to observe the performance on reducing communication costs and to provide a summary on choosing $\lambda$.
% We typically choose a small value for $\lambda$ to avoid over-pre-simplification.

\para{Configurations.} 
We use $16$ CPU-only nodes for the analysis, evaluating performance for $\lambda \in \{0, 1, 10, 10^2, 10^3, 10^4, 10^5\}$ (all smaller than $\Lambda$ for both datasets). 
When $\lambda=0$, there is no pre-simplification, leaving the pipeline unchanged~\cite{LCR24}.

\para{Evaluation metrics.}
To evaluate the runtime and communication workload for attachment points, we consider statistics for three types of communication workload: the highest number of attachment points received by a block, which depends on the pre-simplification thresholding process; the highest number of best up/down volume information received, which tracks the number of supernodes in the largest augmented contour tree of other blocks; the highest number of received branch outer end information, reflecting the highest number of branches in other blocks.

\para{{\Nyx} dataset.}
%%% Logic: we diagnose the best parameter for each dataset using both runtime and workload plots.
We examine parameter sensitivity for the {\Nyx} dataset using subvolumes with $1024^3$ and $2048^3$ voxels.
The statistics for runs on the $2048^3$ subvolume with $\lambda \in \{0, 1\}$ are incomplete because these runs failed midway because the data size  during branch decomposition exceeded the MPI communication size limit.

\cref{fig:parameter-runtime} (top left) shows that runtime on the $1024^3$  {\Nyx} subvolume is stable for $\lambda \geq 10$. 
Similarly, on the $2048^3$ subvolume of {\Nyx} (\cref{fig:parameter-runtime} bottom left), runtime stabilizes after $\lambda=10$.

\cref{fig:parameter-Nyx-communication} shows communication statistics for the {\Nyx} dataset at two subvolume sizes ($1024^3$ and $2048^3$). As the pre-simplification threshold $\lambda$ increases, the number of attachment points exchanged decreases nearly linearly (1st row), while the amount of best up/down information plateaus beyond $\lambda = 100$ (2nd row). The amount of branch information exchanged decreases only slightly, showing minimal sensitivity to $\lambda$ (3rd row).
Among the three metrics, the number of attachment points becomes less important because it is consistently smaller than the number of supernodes, which reflects the amount of best up/down information.

Improvements in both runtime and communication cost fall off after $\lambda=100$, which is consistent across both subvolumes of different sizes.
The consistency arises because the increase in data size primarily reflects a larger observable domain, while the typical feature volume remains relatively stable. Therefore, increasing $\lambda$ for larger subvolumes of the {\Nyx} dataset is unnecessary.

\para{{\microns} dataset.} 
We apply the same evaluations to {\microns}: runtime performance is in the right column of~\cref{fig:parameter-runtime}, and communication cost in~\cref{fig:parameter-microns-communication}.
As with {\Nyx}, we observe the bottleneck effect for the speed and communication cost improvement at $\lambda=100$ for both the $1024^3$ and $2048^3$ subvolumes. 

\para{Summary.}
In summary, we can choose $\lambda\geq 100$ for the optimal performance on both datasets. If users can estimate $\Lambda$ (the volume of the smallest preserved feature), they can choose $\lambda$ with any value in $[100, \Lambda)$. Otherwise, without knowledge of the data to determine $\Lambda$, one would choose a small $\lambda$. Unless otherwise specified, for the remaining experiments on both datasets, we choose $\lambda=100$, which is very small for the domain of size $1024^3$ or larger. 
In addition, we provide a discussion for choosing $\lambda$ in the appendix.

\section{Results and Evaluation } 
\label{sec:results}

In this section, we give experimental results and evaluate scalability. We start with our largest computed contour tree in~\cref{sec:achievement}, followed by the performance improvement in speed and data size compared to previous work  in~\cref{sec:comparison}, then evaluate the algorithm's scalability in~\cref{sec:scalability} using strong scaling and weak scaling.

\begin{figure*}[!ht]
\centering
\includegraphics[width=1.95\columnwidth]{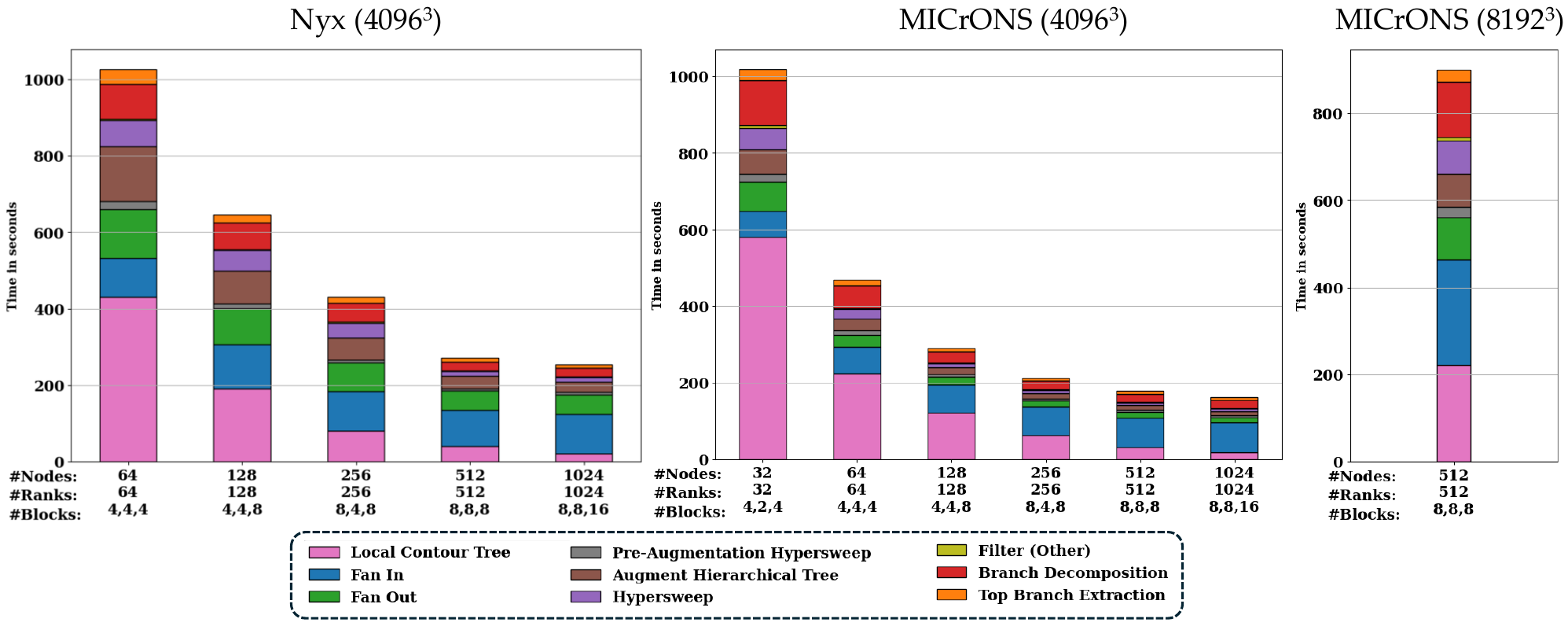}
\vspace{-3mm}
\caption{Runtime performance using OpenMP on Perlmutter for the $4096^3$ volume of {\Nyx} (left),  the $4096^3$ subvolume of {\microns} (middle), and the $8192^3$ volume of {\microns} (right), respectively.}
\label{fig:strong-scaling}
\vspace{-2mm}
\end{figure*}

\subsection{Performance Evaluation}
\label{sec:achievement}
We compute a distributed hierarchical contour tree and its volumetric branch decomposition with pre-simplification on a $8192^3$ subvolume (roughly $550$ billion of voxels) of {\microns}, in less than $15$ minutes; see~\cref{fig:strong-scaling} (right). 
We use $512$ CPU-only nodes and approximately $120$ TiB of memory. 
We apply $\lambda=1000$ for pre-simplification, $10\times$ higher than indicated by the parameter sensitivity analysis.
We produce a contour tree with more than half a trillion regular nodes, which is, to the best of our knowledge, the largest computed contour tree in the literature that supports advanced analytic tasks such as branch decomposition. 

\subsection{Performance Comparison}
\label{sec:comparison}
\para{Existing implementations.}
We compare our distributed computation with pre-simplification to some existing implementations of contour tree computation and volumetric branch decomposition, in particular the standard serial algorithm---Sweep and Merge~\cite{CarrSnoeyinkAxen2003}--- and the state-of-the-art shared-memory parallel implementation---Parallel Peak Pruning (PPP)~\cite{CarrRubelWeber2022a, HristovWeberCarr2020}, as well as the distributed implementations with~\cite{LCR24} and without pre-simplification.

All experiments are conducted on the same type of nodes on the Perlmutter supercomputer; see~\cref{sec:experiment-setting} for configurations. 
Due to memory limits, all shared-memory experiments use the $1024^3$ subvolume of {\microns}. We collect runtime and memory costs, with the number of threads ranging from $1$ to $128$ for the PPP shared-memory implementation. We do not include runtime for top branch extraction (Phase 8, see~\cref{sec:experiment-setting}) in the comparison because shared-memory implementations do not include this step. 

\begin{figure}[!ht]
\centering
\includegraphics[width=1.0\columnwidth]{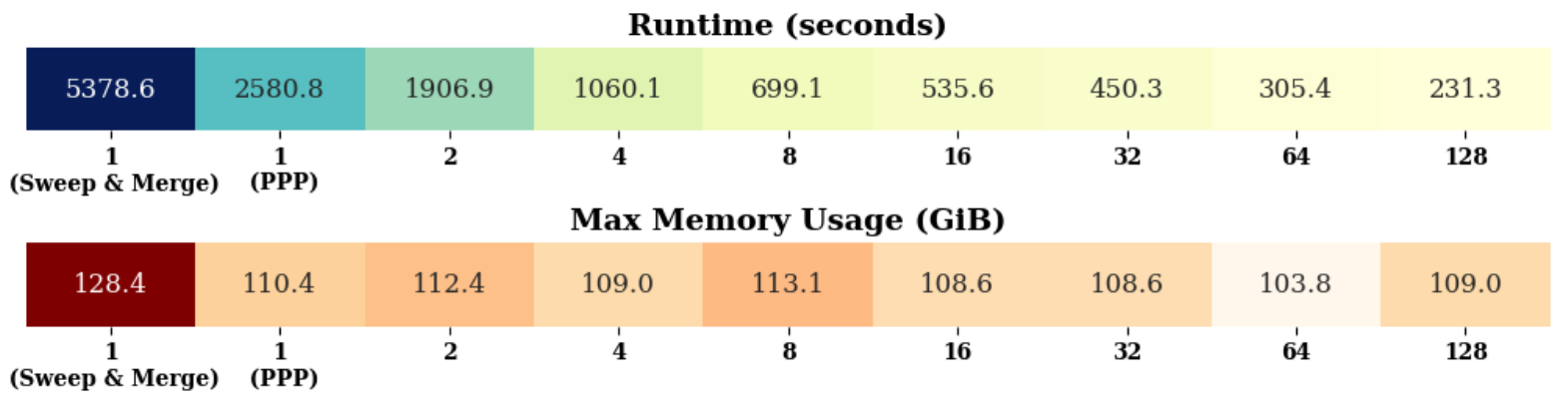}
\vspace{-6mm}
\caption{Performance metrics by number of threads for shared-memory implementations on the $1024^3$ subvolume of {\microns}.} 
\label{fig:single-node-performance}
\vspace{-2mm}
\end{figure}

\para{Runtime performance.}
\cref{fig:single-node-performance} gives performance results for all shared-memory implementations. The PPP implementation in serial is $2.08\times$ faster than the serial implementation due to algorithmic improvements and the internal optimization of VTK-m. 

\begin{figure}[!ht]
\centering
\includegraphics[width=1.0\columnwidth]{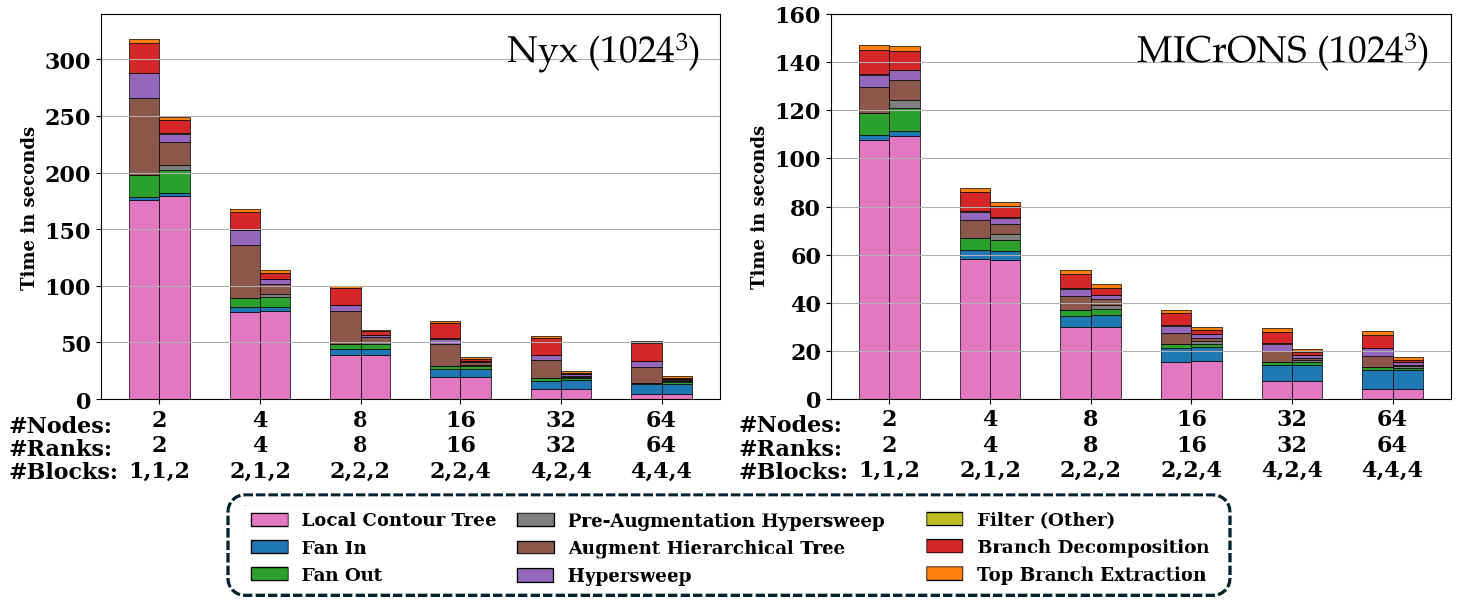}
\vspace{-4mm}
\caption{Grouped bar charts for distributed runtime on $1024^3$ subvolumes of {\Nyx} and {\microns}. The left and right bars of each group show runtime with $\lambda=0$ and $\lambda=100$, respectively.}
\label{fig:runtime-distributed-improvement}
\vspace{-2mm}
\end{figure}

In parallel, however, runtime performance of PPP reaches up to about $11\times$ speedup with $128$ threads compared to serial runs. On top of shared-memory parallelism, the distributed contour tree implementation provides additional improvements. ~\cref{fig:runtime-distributed-improvement} (right) presents the runtime performance of distributed computation without ($\lambda=0$) and with ($\lambda=100$) pre-simplification on the same subvolume of {\microns}. 
Using $64$ nodes, runs without and with pre-simplification spent $26.52$ seconds and $16.09$ seconds, respectively.
Therefore, on the $1024^3$ subvolume of {\microns}, our implementation with pre-simplification reaches an estimated speedup of nearly $334\times$ over the Sweep and Merge serial version and roughly $160\times$ that of the PPP implementation with one thread.
Compared to the parallel run with $128$ threads, our distributed computation with pre-simplification reaches about $14.4\times$ speedup.

While memory limits prevent a single serial run for comparison, we can estimate serial runtime for a $8192^3$ volume, assuming sufficient memory.  We know that 
serial time complexity is dominated by the $O(n \log n)$ term from sorting, and recall the runtime for size $n=2^{30}$ ($1024^3$) from in~\cref{fig:single-node-performance}.
A volume of $2^{39}$ ($8192^3$) voxels is $512\times larger$, with an increased log factor of $39/30 = 1.3$, and we therefore estimate runtime for the Sweep and Merge implementation and the PPP (one thread) implementation to be approximately $3.58 \times 10^6$ seconds and $1.72 \times 10^6$ seconds, respectively. 
Our distributed computation with pre-simplification instead completes in $871.72$ seconds; see~\cref{fig:strong-scaling} right column. 
The estimated speedup of our distributed implementation with pre-simplification is thus up to $4100\times$ over the Sweep and Merge implementation and roughly $1970\times$ that of the PPP implementation in serial.

While Li \etal~\cite{LCR24} showed significant speedup of distributed computation over shared-memory, our pre-simplification process further reduces the communication cost for the distributed analytic computations; see~\cref{sec:parameter-sensitivity}.
We demonstrate runtime improvement on the $1024^3$ subvolumes of {\Nyx} and {\microns}s in~\cref{fig:runtime-distributed-improvement}, in which we compare the runtime of experiments via the grouped bar charts.
For each group of the bar plot in~\cref{fig:runtime-distributed-improvement}, the left bar shows the run without pre-simplification ($\lambda=0$), and the right bar is the runtime with pre-simplification ($\lambda=100$).

We can see the speed improvement on both datasets using pre-simplification under all node/rank configurations. 
Specifically, the runtime for analytical computation phases (all phases after fan-out) is improved by a large margin compared to the runs without pre-simplification. 
For {\Nyx}, the largest total runtime improvement is about $2.58 \times$ using $64$ nodes. 
Similarly, for {\microns}, the largest total runtime improvement is about $1.60\times$.

\begin{figure}[!ht]
\centering
\includegraphics[width=1.0\columnwidth]{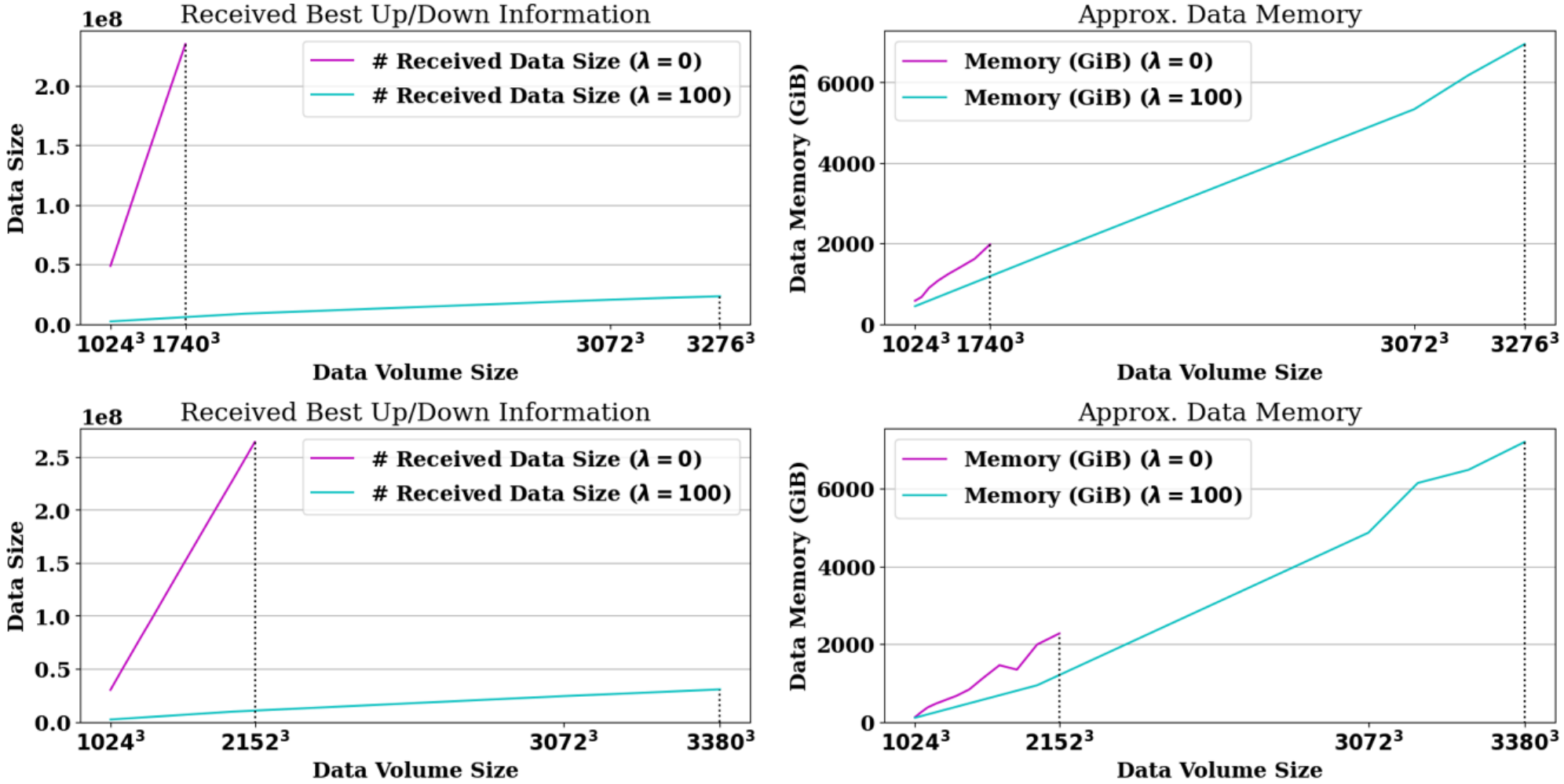}
\vspace{-6mm}
\caption{The left column shows exchanged information size for best up/down volume data of supernodes w.r.t.~data volume size. 
The right column shows approximate total data memory compared to data volume size. Top and bottom rows are for {\Nyx} and {\microns}, respectively. Vertical dotted lines indicate the largest feasible data volume size for the corresponding $\lambda$ choices using $16$ nodes.}
\label{fig:data-size-limit}
\vspace{-4mm}
\end{figure}

\para{Maximum feasible data size.} 
The maximum feasible data size for shared-memory implementations is limited by the available memory per compute node. As shown in~\cref{fig:single-node-performance}, the contour tree for a $1024^3$ subvolume takes $103\sim 129$ GiB of memory. Given a limit of 512 GB ($\approx 476$ GiB) per CPU-only node, linear memory growth would predict an upper limit around $1582^3 \sim 1700^3$ voxels. In contrast, our distributed approach with pre-simplification successfully processes a much larger volume of $8192^3$ voxels, which is difficult to process on existing shared-memory platforms.

% For distributed contour tree computation, there are two bottlenecks limiting the maximum feasible data size: data memory and MPI communication.
While previous work~\cite{CarrRubelWeber2022b, LCR24} significantly increased data scale through distributed computation, this came at the cost of memory overhead due to augmentation, which reduced the maximum data size that can be processed under fixed memory constraints.  
% Besides, our experiments in~\cref{sec:parameter-sensitivity} have shown that the data communication size may exceed the MPI limit when the data size is too large, while no pre-simplification is applied.

We examine maximum data size due to pre-simplification with ($\lambda=100$) and without ($\lambda=0$) pre-simplification, by gradually increasing the subvolume size from $1024^3$ to $4096^3$ in steps of $100$ or $104$ ($\approx 0.1\times1024$, while the boundary size needs to be divisible by $4$) for both datasets. We fix the number of nodes at $16$.
We measure the communication size for the best up/down volume data information during the branch decomposition computation and the total memory consumed by the run reported by Perlmutter.

\cref{fig:data-size-limit} demonstrates the statistics of evaluated metrics for {\Nyx} in the top row and {\microns} in the bottom, in which the vertical dotted lines represent the maximum feasible data size for the two experimental configurations.
Without pre-simplification ($\lambda=0$), the framework can only compute a subvolume of up to $1740^3$ voxels for {\Nyx} and $2152^3$ for {\microns}.
In contrast, as we apply the pre-simplification with $\lambda=100$, the largest feasible data size we can compute grows to $3276^3$ for {\Nyx} and $3380^3$ for {\microns}. 
The maximum feasible volume size increase owing to the pre-simplification is roughly $6.67\times$ for {\Nyx} and $3.87\times$ for {\microns}.

For the runs without pre-simplification, the data communication size grows significantly faster than that of the runs with pre-simplification for our implementation; see~\cref{fig:data-size-limit} left column. 
The runs without pre-simplification crashed after exceeding the one-time MPI communication limit.
For finished runs without pre-simplification, their memory consumption (magenta line) is consistently higher than that of the runs with pre-simplification (cyan line); see~\cref{fig:data-size-limit} right column.

% We analyze the bottlenecks limiting the data size.
% For the runs without pre-simplification, the data communication size grows significantly faster and reaches the MPI communication limit much sooner than that of the runs with pre-simplification; see~\cref{fig:data-size-limit} left column. 
% It indicates that the large number of attachment points is the bottleneck for communication if no pre-simplification is applied.
% In contrast, when the runs with pre-simplification reach the memory capacity (see~\cref{fig:data-size-limit} right column), the data communication size is still far below the MPI limit (\cref{fig:data-size-limit} left column, cyan curves). 
% The MPI communication limit has been largely addressed by the pre-simplification strategy.

Lastly, we discuss memory efficiency by evaluating the footprint per input voxel for all implementations. In a distributed hierarchical contour tree, each block stores a copy of the shared contour tree structure, increasing the total memory footprint.
In addition, augmentation exchanges attachment points across blocks, leading to multiple copies of attachment points in the memory.
Therefore, it is expected that the distributed implementations would have lower memory efficiency than the shared-memory ones.

Following~\cref{fig:single-node-performance}, shared-memory requires approximately $103 \sim 129$ bytes per voxel in the input mesh on {\microns}.
For distributed computations, the memory footprint needed for each voxel is reflected in~\cref{fig:data-size-limit} right column.
On {\microns}, the run without pre-simplification on the $2152^3$ subvolume needs $2273$ GiB; the average memory footprint per input voxel is roughly $244$ bytes.
In contrast, with pre-simplification, the distributed computation on the $3380^3$ subvolume consumes $7199$ GiB; the average memory usage for each input voxel is approximately $200$ bytes.
In other words, while the pre-simplification strategy improves the overall memory efficiency, the extent of the improvement is moderate.
Regardless of pre-simplification, the memory usage for distributed contour tree computation is roughly twice as much as the shared-memory implementations. 
Such sacrifice in memory efficiency is acceptable since we can increase the total memory size by adding compute nodes.

\subsection{Scalability Evaluation}
\label{sec:scalability}

\para{Strong scaling.}
We evaluate strong scaling of the pre-simplified pipeline by measuring performance at a fixed problem size while varying the number of compute nodes.
For this, we use fixed input volumes of size $4096^3$ for both {\Nyx} and {\microns}. The number of nodes is increased from the minimum feasible option for each dataset---$64$ nodes for {\Nyx} and $32$ nodes for {\microns}---up to $1024$ nodes, with one MPI rank assigned per node.  %throughout the experiments.

\cref{fig:strong-scaling} demonstrates the strong scaling plot for the runtime performance of our framework on the $4096^3$ volume of both datasets. The stacked box plots separate the overall runtime by the pipeline phases. 
As we add more nodes, the runtime on both datasets consistently decreases, reaching the near-optimal value at $512$ nodes. 

Among the runtime of phases, all but the fan-in phase have gained a noticeable amount of speedup as the number of nodes increases for both datasets. 
Recall that the analytical computation phases require communication, the size of which is heavily affected by the number of attachment points. 
Without pre-simplification, Li \etal~\cite{LCR24} have shown the scalability limitation of these analytical computation steps due to attachment points (reflected in~\cref{fig:runtime-distributed-improvement}). 
With the pre-simplification strategy, the limitation in scalability is largely mitigated.
While there is still communication overhead for the shared tree structure, the analytical computation steps are no longer the scalability bottleneck.

With pre-simplification, the bottleneck of scalability becomes the fan-in operation, which is a critical step in constructing the contour tree. In this step, shared boundary information needs to be exchanged between adjacent blocks to construct the shared contour tree structure that goes across the block boundary. We are unsure whether this can be optimized, so we leave it for future work.

\begin{figure}[!ht]
\centering
\includegraphics[width=1.0\columnwidth]{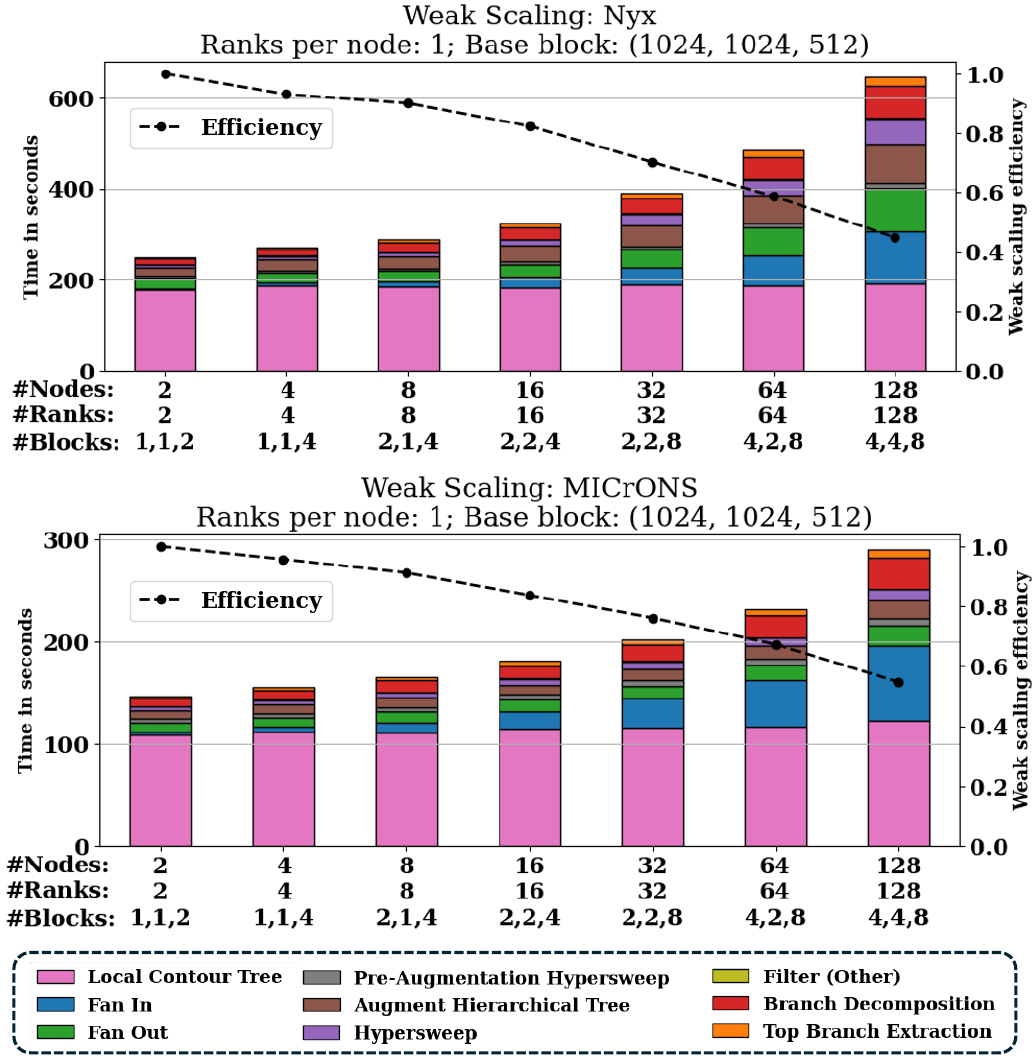}
\vspace{-6mm}
\caption{Runtime performance with a growing number of nodes (and blocks) and data size, while each node is assigned
a $1024\times1024\times 512$ subvolume (weak scaling) with one MPI rank per node. The top and bottom rows are for {\Nyx} and {\microns}, respectively.}
\label{fig:weak-scaling}
\vspace{-6mm}
\end{figure}

\para{Weak scaling.}
We provide the weak scaling analysis for the optimized framework.
For both {\Nyx} and {\microns}, we start with a subvolume of $1024\times 1024\times 512$ voxels. 
As we increase the number of nodes and blocks, we simultaneously grow the subvolume size accordingly so that each compute node is assigned a subvolume at a fixed size.

We report the runtime and the weak scaling efficiency in~\cref{fig:weak-scaling}.
Among all the phases, only the local contour tree computation has a roughly constant runtime. 
All other phases become increasingly expensive as the number of nodes grows.
This is because the amount of work to communicate and process the data for the shared tree structure naturally increases, leading to increasing communication overheads and a drop in the weak scaling efficiency~\cite{CarrRubelWeber2022b}. 
On the other hand, the runtime growth for all the analytical computation phases is comparable to or slower than that of the fan-in phase. 
In other words, with our pre-simplification strategy, such analytical computations are not the bottlenecks of the algorithm's scalability.

\section{Conclusion}
\label{sec:conclusion}

We introduce a pre-simplification strategy to optimize the analytical computation for distributed hierarchical contour trees. 
First, with the pre-simplification process, we generate the largest known contour tree on a volume of size $8192^3$ with complex topology within $15$ minutes.
Second, we demonstrate a runtime speedup of up to $334\times$ over the serial computation and $14.4\times$ over the parallel implementation on a $1024^3$ dataset.
Assuming there is enough memory for the serial computation on the data volume of size $8192^3$, our performance is expected to reach up to $4100\times$ speedup over the serial version in theory.
Moreover, our pre-simplification strategy enables $1.60 \sim 2.58 \times$ speedup for the distributed computation and supports $3.87 \sim 6.67\times$ larger size of data volume with a fixed number of compute nodes. 
Lastly, our pre-simplification strategy has largely mitigated the scalability issue of the distributed contour tree computations in~\cite{LCR24}.

%We intend to explore other measures of contour trees besides volume for enhanced utility and flexibility of the distributed augmented contour tree for various scientific applications.

\acknowledgments{
This research was supported by the U.S. Department of Energy (DOE), Office of Science, Advanced Scientific Computing Research (ASCR) program and the Exascale Computing Project (17-SC-20-SC), a collaborative effort of the DOE Office of Science and the National Nuclear Security Administration under Contract No.~DE-AC02-05CH11231 to the Lawrence Berkeley National Laboratory.
This research used resources of the National Energy Research Scientific Computing Center (NERSC), a Department of Energy Office of Science User Facility using NERSC award ASCR-ERCAP0026937.
Additionally, Mingzhe Li and Bei Wang were partially supported by DOE DE-SC0021015 and National Science Foundation (NSF) IIS-2145499. Hamish Carr was supported by the University of Leeds.
}

\bibliographystyle{abbrv-doi}

\bibliography{refs-presimplify}

\clearpage
\appendix
\section{Discussion: Parameter Estimation}
\label{sec:discussion-parameter}

In~\cref{sec:parameter-sensitivity}, we conducted a parameter sensitivity analysis on the two datasets to determine $\lambda$, running a sequence of exponentially increasing $\lambda$ values. This analysis provides a systematic approach for identifying the optimal $\lambda$. In this section, we further discuss criteria to help users estimate an appropriate value for $\lambda$, focusing in particular on determining its minimum value.

\para{Estimating the minimum value of $\lambda$.}
We begin by defining the terms used in the discussion. Let $N$ denote the total data volume and $R$ the number of ranks. The parameter $\lambda$ determines which attachment points are exchanged: specifically, only those whose interior forest volume exceeds $\lambda$ are included. Let $\alpha$ be the number of attachment points received from other ranks. An upper bound on $\alpha$ for each rank is 
\[
\frac{N - N/R}{\lambda + 1}, 
\]
where $N - N/R$ represents the maximum external data volume for a rank, and $\lambda + 1$ specifies the minimum interior forest volume of attachment points to be exchanged. This bound is loose, as part of the data volume is already represented in the shared contour tree structure.

The first criterion for determining $\lambda$, referred to as the \emph{memory criterion}, ensures that sufficient memory is available for analytical computations. Based on this criterion, we estimate the minimum value of $\lambda$ through two runs on a subvolume of the data, followed by a test run on the full dataset.

For example, to estimate the minimum $\lambda$ satisfying the memory criterion for the $2048^3$ volume of the {\Nyx} dataset, we first run the framework on a $1024^3$ subvolume with two values of $\lambda$: $0$ and $100$. The run without pre-simplification ($\lambda = 0$) consumes $574.66$ GiB of memory, whereas the run with $\lambda = 100$ uses $439.17$ GiB. In parallel, the number of attachment points drops from $697{,}320{,}285$ to $1{,}288{,}810$. This reduction implies that processing approximately $6.96 \times 10^8$ attachment points requires about $135.49$ GiB of memory, or roughly $209.02$ bytes per attachment point.

Next, we perform a test run on the full $2048^3$ volume using a large $\lambda$ value (e.g., $\lambda = 10^5$) to eliminate most attachment points and measure memory consumption. If this run fails due to insufficient memory, the hardware configuration is unlikely to be viable for the dataset, regardless of $\lambda$. If successful, we record the peak memory usage for any single rank—in our case, $133.26$ GiB (note that this is not the total memory usage across all ranks). The remaining available memory must then be sufficient to handle attachment point computations.

In this example, $N = 2048^3$, $R = 16$, and each rank (i.e., compute node) has $512$ GB of available memory. Since the upper bound on the number of attachment points is
$
\frac{N - N/R}{\lambda + 1},
$
and the memory required for attachment point computation is $209.02$ bytes per voxel on average, we require
\[
209.02\,\alpha < 209.02 \frac{N - N/R}{\lambda + 1} < 512 \times 10^9 - 133.26 \times 1024^3.
\]
From this, we estimate the minimum $\lambda$ for this example to be $4$.

The second criterion concerns communication overhead, which we refer to as the \emph{communication criterion}. Pre-simplification reduces the number of attachment points exchanged during communication, thereby mitigating scalability limits. As shown in \cref{sec:parameter-sensitivity}, both the attachment points and the shared contour tree structure contribute to this overhead. With increasing $\lambda$, the number of attachment points decreases, and eventually the shared contour tree structure becomes the dominant factor. For optimal scalability, our goal is to reduce the number of attachment points to be comparable to, or smaller than, the size of the shared contour tree, which has been shown~\cite{HristovWeberCarr2020, CarrRubelWeber2022b, LCR24} to be bounded by $O(N^{2/3})$ for 3D data. This implies that $\lambda$ should be on the order of $\Omega(N^{1/3})$ for optimal scalability.

\para{Limitation.}
We conclude by discussing the limitations of our approach for estimating $\lambda$. First, the memory criterion requires recording statistics such as the number of attachment points and the memory usage for each rank. Although our implementation includes logging functionality, this method still entails additional effort. Second, the estimated minimum $\lambda$ for the first criterion is likely higher than the true minimum, as it is based on the worst-case distribution of attachment points. Third, for the communication criterion, constant factors in the computation make it difficult to determine a precise minimum value of $\lambda$.

\para{Parameter choices.}
While pre-simplification significantly reduces communication overhead and enables the processing of much larger datasets, it also removes some small-volume features that may be important in certain tasks or data contexts. To preserve such features, we aim to choose $\lambda$ as small as possible, subject to satisfying the memory criterion (and optionally the communication criterion), and ensuring that $\lambda < \Lambda$. However, if the estimated $\lambda$ is substantially larger than the expected size of the smallest relevant features, pre-simplification may not be suitable for the application.

Currently, our implementation supports only a single, global $\lambda$ for pre-simplification as a means of reducing the number of attachment points in the computation. Since the augmentation step can be performed on arbitrary subsets of attachment points~\cite{LCR24}, it would be possible to customize $\lambda$ for different subareas or subvolumes of the data, depending on the features of interest, or even to selectively preserve specific features—an extension we leave for future work.

\end{document}